# Spintwistronics: Photonic bilayer topological lattices tuning extreme spin-orbit interactions

Peng Shi[1,†,*], Xinxin Gou[1,†], Qiang Zhang[1,†,*], Weiyu Wei[1], Haijun Wu[2,4], Songze Li[1], Zhihan Zhu[4], Yijie Shen[2,3*] and Xiaocong Yuan[1,5,*]

[1] *Nanophotonics Research Center, Shenzhen Key Laboratory of Micro-Scale Optical Information Technology & Institute of Microscale Optoelectronics, Shenzhen University, Shenzhen 518060, China*

[2] *Centre for Disruptive Photonic Technologies, School of Physical and Mathematical Sciences, Nanyang Technological University, Singapore 637371, Singapore*

[3] *School of Electrical and Electronic Engineering, Nanyang Technological University, Singapore 639798, Singapore*

[4] *Wang Da-Heng Center, HLJ Key Laboratory of Quantum Control, Harbin University of Science and Technology, Harbin 150080, China*

[5] *Zhejiang Lab, Research Center for Humanoid Sensing, Research Institute of Intelligent Sensing, Hangzhou 311100, China*

[†]*These authors contributed equally to this work*
*\*Corresponding authors*

E-mail: pittshiustc@gmail.com, zhangqiang@szu.edu.cn, yijie.shen@ntu.edu.sg and xcyuan@szu.edu.cn

Twistronics, the manipulation of Moiré superlattices via the twisting of two layers of two-dimensional (2D) materials to control diverse and nontrivial properties, has recently revolutionized the condensed matter and materials physics[1-6]. Here, we introduce the principles of twistronics to spin photonics, coining this emerging field spintwistronics. In spintwistronics, instead of 2D materials, the two layers consist of photonic topological spin lattices on a surface plasmonic polariton (SPP) platform. Each 2D SPP wave supports the construction of topological lattices formed by photonic spins with stable skyrmion topology governed by rotational symmetry. By introducing spintwistronics into plasmonics, we demonstrate theoretically and experimentally that two layers of photonic spin lattices can produce Moiré spin superlattices at specific magic angles. These superlattices, modulated periodically by the quantum number of total angular momentum, exhibit novel properties-including new spin quasiparticle topologies, multiple fractal patterns, extremely slow-light control, and more-that cannot be achieved in conventional plasmonic systems. As a result, they open up multiple degrees of freedom for practical applications in quantum information, optical data storage and chiral light-matter interactions.

## Introduction

When two layers of identical graphene are superposed with a controlled twist angle between them, Moiré superlattices can form, remarkably giving rise to a range of nontrivial electronic[1]. For example, at a magic angle of approximately 1.08°, flat bands appear around the electronic Fermi energy levels, leading to a correlated insulating phase[2]. Shortly thereafter, other magic angles were found to induce additional exotic properties within bilayer graphene's Moiré lattices, such as unconventional superconductivity[3], higher-order topological insulating states[4], tunable spin-polarized phases[5], and Chern insulators[6], etc. This approach has catalyzed the rapid development of twistronics, in which the simple superposition of two layers of 2D materials can yield new Moiré periodic structures with fundamentally altered properties compared to the original monolayer materials. Twistronics has since expanded to other materials, including various van der Waals optoelectronic materials[7], mechanical systems[8], acoustic lattices[9], and magnetic skyrmion materials[10], thereby transforming multiple domains within physics and materials science.

Twistronics has recently entered the optical domain, and begun to revolutionize photonic technologies[11]. By generating Moiré patterns in optical lattices, it is now possible to excite flat bands within laser-written materials, enabling control over the (de)localization of light fields[12,13], optical soliton formation[14,15], skyrmion bags topology[16,17], Thouless pumping[18], and reconfigurable moiré nanolaser[19,20]. Twisting two layers of photonic materials with van der Waals coupling has proven effective in exciting topological polaritons[21-23]. Additionally, twisting photonic crystal slabs has been used to control flat-band phenomena[24,25], and generate vortex beams[26]. Metasurface twisting has recently emerged as a powerful tool for beam steering[27] and non-diffraction control[28]. Extending twistronics into various photonic fields holds exciting potential for producing novel physical effects.

In this Letter, we introduce the concept of twistronics applied to topological spin lattices on surface plasmon polariton (SPP) waves[29-32], creating a new fusion of twistronics and topological spin photonics. SPP waves have recently led to the development of various forms of topological quasiparticles[16,17,29-33]. Additionally, optical spin lattices can be constructed on SPP waves with periodic textures, enabling the formation of a series of skyrmion and meron topologies through symmetry control[34-36]. By incorporating twistronics, we demonstrate that a bilayer of photonic spin lattices can produce diverse Moiré spin superlattices at specific magic angles, resulting in novel properties, such as unique quasiparticle topologies, fractal patterns, slow-light effects, etc., that are unattainable through conventional plasmonics. Moreover, in contrast to recent consideration of twisted electric field SPP waves[16,17], our concept includes spin-orbit couplings, providing extra degree-of-freedom to control the light-matter interactions, offering applications in the quantum information, optical data storage and chiral light-matter interactions.

## Photonic Moiré spin superlattice

Conventional photonic topological spin textures can be realized on a transverse magnetic (TM) SPP platform, where diverse topological lattices are engineered by adjusting the rotational symmetry of the optical system. For example, lattices with $C_6$-symmetry yield fractal, skyrmion-like topological spin textures, whereas $C_4$ or $C_3$-symmetries generate meron-like spin textures[34-36], as shown in the top panels of **FIG. 1**(a) and **1**(c). Moiré superlattices are then formed by superimposing two identical layers of these topological spin lattices with a controllable twist angle, illustrated in the bottom panels of **FIG. 1**(a) and **1**(c). For instance, at a twist angle of $2\vartheta_{C4} = \arctan(3/4)$ and a total angular momentum (TAM) quantum number $l = 7$, the Moiré superlattices of $C_4$-symmetric sublattices exhibits a skyrmion configuration with a topological skyrmion number $n_{sk} = \pm 1$. Conversely, when the twist angle is $2\vartheta_{C4} = \arctan(8/15)$ and the TAM quantum number is $l = 2, 4, 6, 10$, meron-like geometric cluster spin textures emerge within the $C_4$ Moiré superlattices. This method enables the construction of a variety of new topological quasiparticles in Moiré superlattices. Distinct from previously studied Moiré lattices of electromagnetic (EM) fields, the topological spin textures in this approach can be finely tuned by the TAM quantum number, which we will explore in detail in our work.

As described above, photonic Moiré spin lattices are created by superimposing topological sublattices, such as skyrmion or meron-like lattices, with an inclination angle of $\pm\vartheta$, as shown in the top panels of **FIG. 1**(a) and **1**(c). For the SPP mode considered in this study, the analytical expression for the electric field component in the normal direction at the air/metal interface ($E_z$) can be derived from the Hertz potential $\Psi$, expressed as[34,35]:

$$\Psi = E_z(+\vartheta) + E_z(-\vartheta) \quad (1)$$

with

$$E_z(\pm\vartheta) = A\sum_{n=1}^{N} e^{il(\theta_n \pm \vartheta)} e^{i\beta \mathbf{r}_\perp(\pm\vartheta)\cdot\mathbf{e}_n} e^{-k_z z}, \quad (2)$$

where $\theta_n = 2n\pi/N$, $\mathbf{e}_n = (\cos\theta_n, \sin\theta_n)$ with $n = 1, \ldots, N$, where $N$ represents the $N$-fold rotational symmetry of the sublattice (denoted as $C_N$). Here, $\beta$ and $ik_z$ are the wavenumbers in the horizontal (propagation constant) and vertical (evanescent wavenumber) directions, respectively; $\beta^2 - k_z^2 = k^2$, with $k$ as the wavenumber in the air half-space. The horizontal position vector $r_\perp(\pm\vartheta) = \rho(\cos(\varphi \pm \vartheta), \sin(\varphi \pm \vartheta))$ is defined with $\rho$ and $\varphi$ representing the radial and azimuthal coordinates in the cylindrical coordinate system, and $i$ is the imaginary unit. The spin angular momenta (SAMs) **S** of these SPP fields in a nondispersive medium can be derived as[37]

$$\mathbf{S} = \frac{\varepsilon\beta^2}{4\omega}\text{Im}(\nabla\Psi^* \times \nabla\Psi) = -\frac{\varepsilon\beta^2}{4\omega}\langle\nabla\Psi|\times i|\nabla\Psi\rangle. \quad (3)$$

The SAM of a SPP wave has a form of the Berry curvature of the Hertz potential[37], and consequently the nontrivial properties of Moiré superlattices are a consequence of spin-orbit coupling from the perspective of Hertz potential. Here, $\omega$ is the angular frequency of the monochromatic time-harmonic EM field, and $\varepsilon$ represents the permittivity of air. The spin textures of sublattices arranged in threefold ($C_3$), fourfold ($C_4$) and sixfold ($C_6$) rotational symmetries are provided in **FIG. S1** to **FIG. S3** in the supplemental materials.

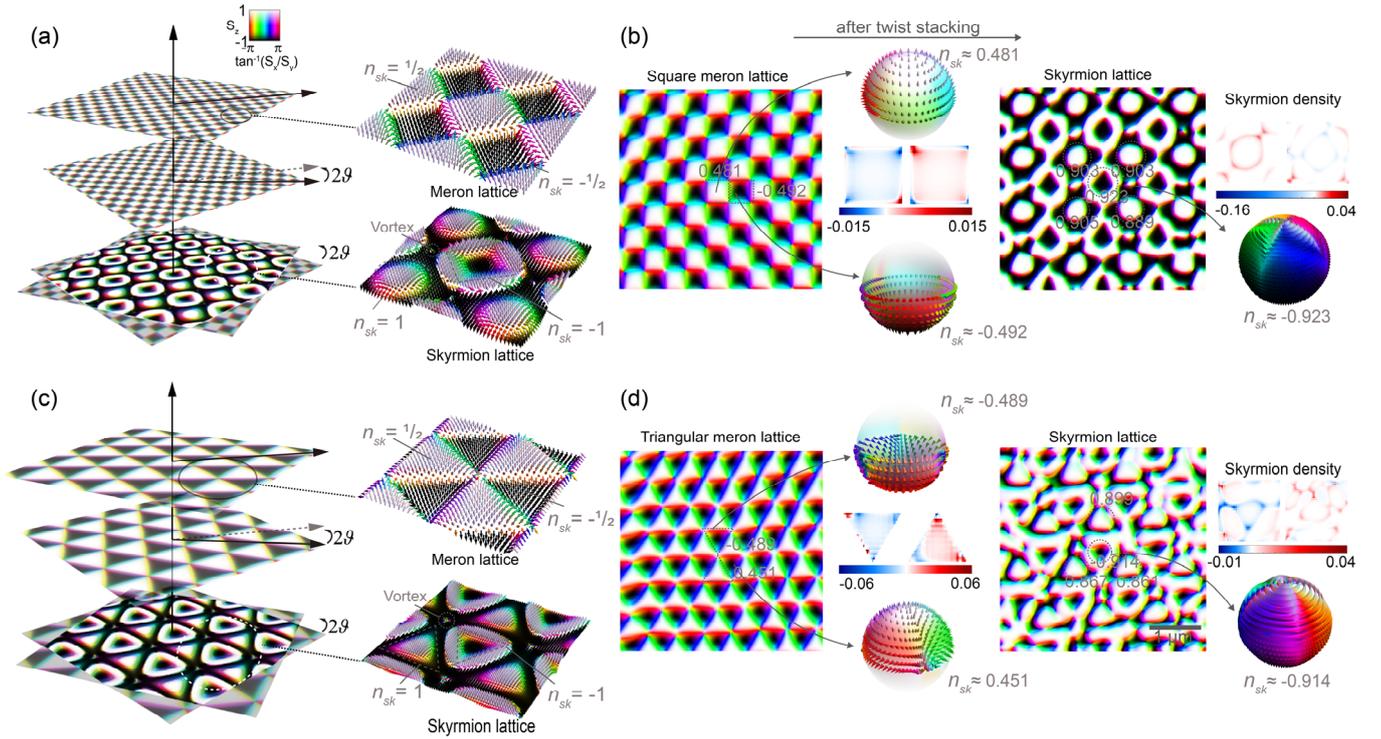

**FIG. 1. Concept of photonic Moiré spin superlattice formation.** (a) A photonic Moiré spin lattice (bottom panel, with a vector diagram of the Moiré lattice shown in the right panel) is formed by superimposing $C_4$-symmetric sublattices (top panel, with a vector diagram of the meron-like texture in $C_4$-symmetry shown in the right panel) at a twist angle $2\vartheta_{C4} = \arctan(3/4)$ and with a TAM quantum number $l = 7$. (b) From the reconstructed spin textures with experimentally measured $S_z$, stereographic projections, and skyrmion number densities (right panel), we observe that each unit cell of the Moiré spin lattice, in $C_4$-symmetry, can be considered a combination of skyrmion topologies with skyrmion number $n_{sk} = \pm 1$. (c) Similarly, a photonic Moiré spin lattice (bottom panel, with a vector diagram of the Moiré lattice shown) is generated by $C_3$-symmetric sublattices (top panel, with a vector diagram of meron-like texture in $C_3$-symmetry shown) at a twist angle $2\vartheta_{C3} = \arccos(1/7)$ and a TAM quantum number $l = 8$. (d) The reconstructed spin textures with experimentally measured $S_z$, along with stereographic projections and skyrmion number densities (right panel), reveal that each unit cell of this Moiré spin lattice in $C_3$-symmetry also forms skyrmion topologies with skyrmion number $n_{sk} = \pm 1$. The spin textures and skyrmion number densities are displayed in the left panels.

From Eq. (2), it can be observed that the Moiré spin lattices are influenced by the quantum number of TAM, in contrast to conventional Moiré lattices of the electric field[12-20]. Consequently, the conditions for the formation of Moiré spin lattices are significantly stricter compared to those for conventional Moiré lattices. For instance, when forming Moiré spin lattices from $C_4$-symmetric sublattices, the twisted angle must satisfy the following relation:

$$2\vartheta_{C4} = \arctan \frac{2(2m_1+1)(2m_2+1)}{\left[(2m_1+1)+(2m_2+1)\right]\left[2m_1-2m_2\right]}, \quad (4)$$

where $m_1$ and $m_2$ can be any integers. From Eq. (4), it can be noted that the denominator is an even integer, while the numerator is an odd integer. For example, when $m_1 = 1$ and $m_2 = 0$, we find that $2\vartheta_{C4} = \arctan(3/4)$. This specific twisted angle results in the formation of the Moiré spin lattice, as illustrated in the right panel of **FIG. 1**(a).

In the case of sublattices exhibiting $C_3$ and $C_6$-symmetries, the twisted angles must satisfy the following conditions:

$$2\theta_{C3} = \arccos\frac{m_1^2 - 3m_2^2}{m_1^2 + 3m_2^2}, \tag{5}$$

and

$$2\theta_{C6} = \arccos\frac{(m_1 - m_2)^2 - 3(m_1 + m_2)^2}{(m_1 - m_2)^2 + 3(m_1 + m_2)^2}, \tag{6}$$

respectively. In fact, these two conditions are coincident. For example, from Eq. (5), when $m_1 = 2$ and $m_2 = 1$, we obtain that $2\vartheta_{C3} = \arccos(1/7)$. This specific twisted angle ensures the formation of the Moiré spin lattice, as shown in the right panels of **FIG. 1**(c).

As discussed earlier, the additional degree-of-freedom associated with the quantum number of TAM imposes an extra restriction on the formation of Moiré spin lattices. However, this also introduces new features and interesting phenomena within the Moiré spin lattices.

The first feature of the Moiré spin lattices is their periodicity with respect to the quantum number of TAM $l$. For Moiré spin lattices constructed from sublattices exhibiting $C_N$ rotational symmetry (where $N = 3, 4, 6$) and a twisted angle $2\vartheta_{CN}$, the period $p$ is determined by the conditions $p \bmod N = 0$ and $p = m\pi/\vartheta$, where $m$ is an arbitrary integer. Generally speaking, the quantity $m\pi/\vartheta$ is not strictly an integer. However, if $p = m\pi/\vartheta$ can be approximated as an integer, it represents the period of the photonic Moiré lattices in relation to the quantum number of TAM. For example, for photonic Moiré lattices formed by $C_4$-symmetric sublattices with a Moiré angle of $2\vartheta_{C4} = \arctan(5/12)$, we find that $p = m\pi/\vartheta \approx 15.915 \approx 16$ when $m = 1$, which obviously satisfies $16 \bmod 4 = 0$. Thus, $p = 16$ can be considered the period of these photonic Moiré lattices, as shown in **FIG. S4** and **FIG. S7**. Similarly, for photonic Moiré lattices constructed from $C_6$-symmetric sublattices with a Moiré angle of $2\vartheta_{C6} = \arccos(11/14)$, we find that $p = m\pi/\vartheta \approx 65.946 \approx 66$ when $m = 7$, satisfying $66 \bmod 6 = 0$. Therefore, $p = 66$ can be regarded as the period of these photonic Moiré lattices, as depicted in **FIG. S5** and **FIG. S8**. A similar condition applies to the Moiré lattices formed by $C_3$-symmetric sublattices, as illustrated in **FIG. S6** and **FIG. S9**.

The second intriguing phenomenon is the formation of Moiré spin superlattices, where the unit cells comprise a combination of topological skyrmions with skyrmion number $n_{sk} = \pm 1$ generated by meron-like sublattices. Previously, only meron lattices could be formed under $C_3$ or $C_4$ rotational symmetries, while fractal skyrmion lattices were associated with $C_6$-symmetry[34,35]. We refer to the spin lattices in $C_6$ rotational symmetry as fractal skyrmion lattices because no regions exists where the integral of the skyrmion number density yields a nonzero integer, indicating that the topology of the SAM is nontrivial only when considering specific extracted sets of these fractal lattices[34,35].

Moiré superlattices introduce additional degrees-of-freedom for constructing novel types of topological lattices, such as twisted angles and the quantum number of TAM. For example, for the Moiré lattices formed by $C_4$-symmetric sublattices with a twisted angle $2\vartheta_{C4} = \arctan(3/4)$ and a quantum number of TAM $l = 7$, the unit cell can be viewed as a combination of skyrmions with skyrmion number[33] $n_{sk} = \pm 1$ in $C_4$-symmetry, as shown in **FIG. 1**(a). In **FIG. 1**(b), we present the reconstructed spin texture alongside the experimentally measured distribution of the $S_z$ component of SAM for both the sublattice (left panel) and the Moiré lattice (right panel) at the twisted angle $2\vartheta_{C4} = \arctan(3/4)$. The inset images display the stereographic projections and the skyrmion number densities for both the sublattices and the superlattices. The calculated skyrmion numbers for the unit cells of experimental sublattices are 0.481 and −0.492, reflecting the presence of meron topologies. In contrast, the calculated skyrmion numbers of the unit cells of the experimental superlattices are −0.923 and 0.903, indicating skyrmion topologies.

Similarly, for the Moiré spin lattices constructed from $C_3$-symmetric sublattices with a twisted angle $2\vartheta_{C3} = \arccos(1/7)$ and a quantum number of TAM $l = 8$, the unit cell can also be regarded as a combination of skyrmions with skyrmion number $n_{sk} = \pm 1$ in $C_3$-symmetry, as shown in **FIG. 1**(c). In **FIG. 1**(d), we display the reconstructed spin textures and the experimentally measured distributions of $S_z$ for the sublattice (left panel) and the Moiré lattice (right panel) at the twisted angle $2\vartheta_{C3} = \arccos(1/7)$, respectively. The inset images reveal the stereographic projections and the skyrmion number densities for both the sublattices and the superlattices. The calculated skyrmion numbers for the unit cells of the experimental sublattices are 0.451 and −0.489, indicative of meron topologies. In comparison, the calculated skyrmion numbers for the unit cells of the experimental superlattices are −0.914 and 0.867, confirming the presence of skyrmion topologies.

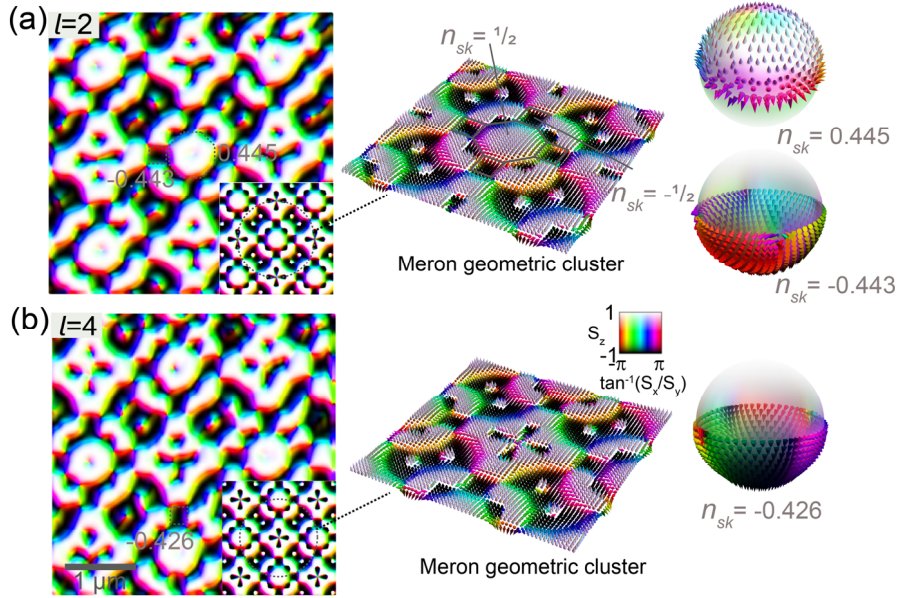

**FIG. 2. Meron geometric cluster-like spin textures in photonic Moiré spin lattices.** In the photonic Moiré spin lattices formed by $C_4$-symmetric sublattices, with a twisted angle of $2\vartheta_{C4} = \arctan(8/15)$, we investigate the meron geometric cluster-like spin textures at different quantum numbers of TAM: (a) $l = 2$ and (b) $l = 4$. The meron geometric cluster is conceptualized as a collection of meron topologies arranged in arbitrary geometries. The reconstructed spin textures, along with the experimentally measured $S_z$ components, are displayed in the left panels, while the corresponding vector diagrams are shown in the middle panels. The integral skyrmion number densities of the meron topologies are found to be −0.443 and 0.445 in (a) and -0.426 in (b), respectively, as evidenced by the stereographic projections in the right panels.

Additionally, meron geometric cluster-like spin textures can be formed within the Moiré spin lattices. A meron geometric cluster can be described as a collection of meron topologies arranged in arbitrary geometries. By tuning the quantum number *l* of the TAM, the geometric arrangement of these meron topologies can be controlled. For example, in the Moiré lattices constructed from $C_4$-symmetric sublattices with a twisted angle $2\vartheta_{C4} = \arctan(8/15)$ and TAM quantum number *l* = 2, 4, 8, and 10, the meron geometric cluster -like textures can be generated. As the quantum number varies from *l* = 2 to *l* = 4, the orientations of the photonic meron topologies can be effectively tuned, as illustrated in **FIG**. 2(a-b). The left panels present the reconstructed spin textures along with the experimentally measured distributions of $S_z$, with inset images showing the corresponding theoretical comparisons. The middle panels display vector diagrams of the meron textures. From the stereographic projections in the right panel, we observe that the calculated skyrmion numbers for the unit cells of the experimental superlattices are −0.443 and 0.445, respectively, indicating the presence of meron topologies. Detailed measurements for the quantum numbers of TAM *l* = 2, 4, 6 and 10 can be found in **FIG. S**16 to **FIG. S**19.

## Fractal spin textures

Fractal (self-similarity) properties are prevalent in the photonic Moiré topological spin lattices. For instance, consider the Moiré superlattices constructed from $C_3$ rotational symmetric sublattices, as shown in **FIG. 3**(a). The $S_z$ component can be expressed as follows:

$$S_z \propto \left\{ \begin{aligned} &+\sin\left[\sqrt{3}\beta(x\sin\vartheta - y\cos\vartheta) - l\frac{4\pi}{3}\right] \\ &-\sin\left[\sqrt{3}\beta\left(x\sin\left[\vartheta + \frac{\pi}{3}\right] - y\cos\left[\vartheta + \frac{\pi}{3}\right]\right) - l\frac{2\pi}{3}\right] \\ &-\sin\left[\sqrt{3}\beta\left(x\sin\left[\vartheta - \frac{\pi}{3}\right] - y\cos\left[\vartheta - \frac{\pi}{3}\right]\right) - l\frac{2\pi}{3}\right] \end{aligned} \right\}. \quad (7)$$

For a twisted angle of $2\vartheta_{C3} = \arccos(-1/26)$ and a quantum number of TAM *l* = 10, the experimentally measured $S_z$ component (with the corresponding theoretical comparison shown in the inset) can be found in **FIG**. 3(b). To investigate the fractal properties of the Moiré superlattices, we perform a Fourier transformation on the experimental SAM density $S_z$. This analysis reveals four sets of wavenumbers in Fourier space, as illustrated in **FIG. 3**(c). By extracting these distinct groups of wavenumbers and performing an inverse Fourier transformation, we can obtain three sets of triangular lattices, presented in **FIG**. 3(d), **FIG**. 3(e) and **FIG**. 3(i), respectively. The left panel in **FIG**. 3(f) contains a pair of sublattices, a result of the Moiré properties of the lattice. By carefully selecting the angles of inclination, we can extract the two sublattices that satisfy $C_3$-symmetry, shown in **FIG**. 3(g) and **FIG**. 3(h).

Both sublattices are observed to be inclined, confirming our earlier analysis. Furthermore, all lattices depicted in **FIG**. 3(c-h) exhibit $C_3$-symmetry, underscoring the fractal properties of the Moiré lattices. Notably, in optical systems, the Moiré lattices are inherently limited by the diffraction limit. Thus, the maximum wavenumber cannot exceed $2k$, where $k$ is the wavenumber in free space. This means that the scale of photonic Moiré fractal lattices cannot be made indefinitely small, highlighting a significant distinction between optical phenomena and natural fractals.

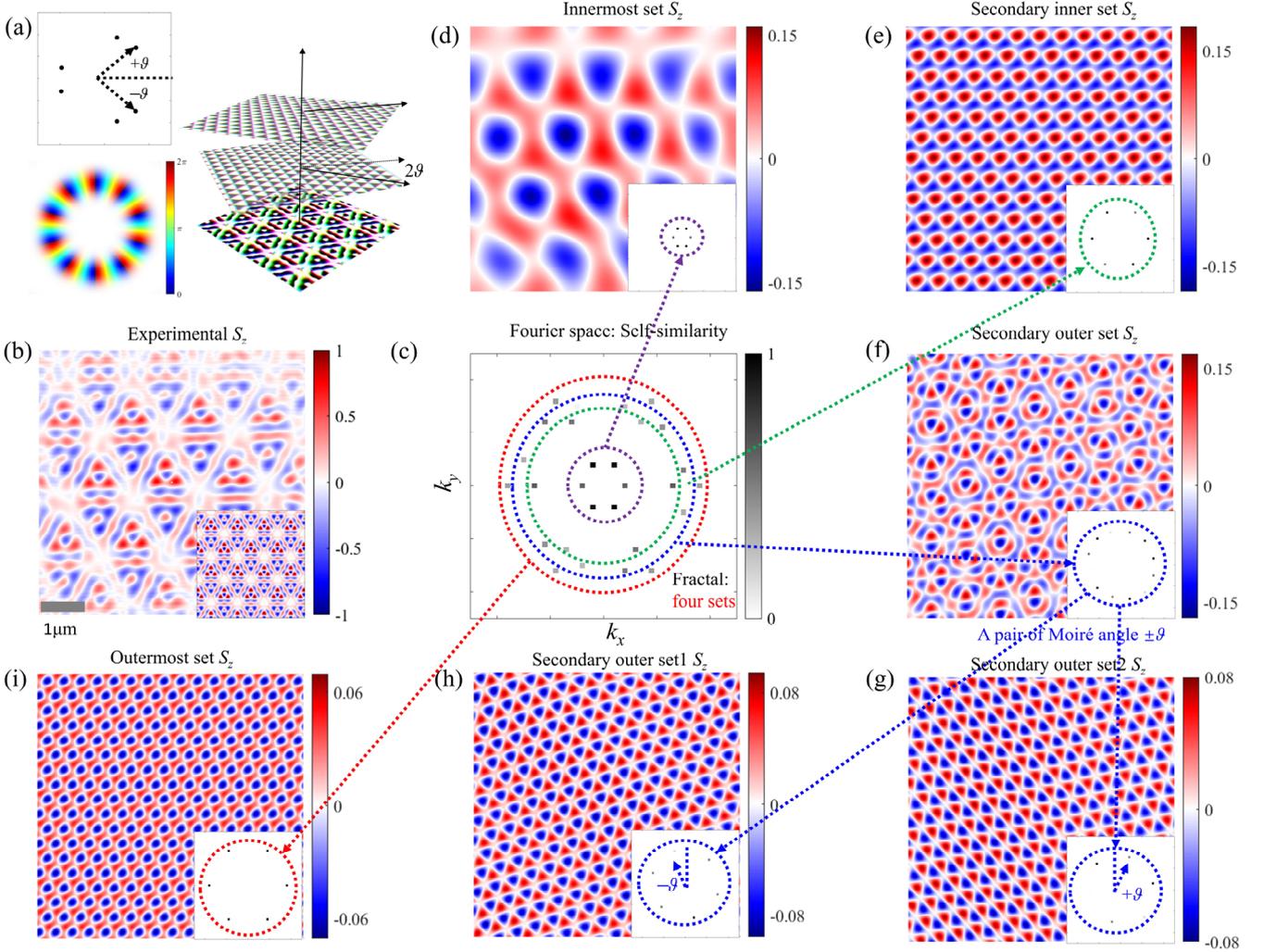

**FIG. 3. Fractals in photonic Moiré lattices**. As the twisted angle $2\vartheta_{C3} = \arccos(-1/26)$ and the quantum number of TAM is $l = 10$, (a) the Moiré lattice is successfully formed. From the experimental measurement of (b) $S_z$ (with the inset image providing a theoretical comparison) and the corresponding (c) wavenumber representation in Fourier space, we observe four distinct sets of wavenumbers. The extracted sets include: (d) the central set located within the purple dashed circle, (e) the second set situated between the purple and green dashed circles, (f) the third set found between the green and blue dashed circles and (i) the fourth set positioned between the blue and red dashed circles. Additionally, the extracted third set can be further divided into two sublattices, as illustrated in (g) and (h). The inset images exhibit the wavenumbers in Fourier space corresponding to each set. Detailed theoretical results can be referenced in **FIG. S**11. The scale is indicated by the grey line in (a).

# Slow light effect

Although there is no localization in photonic Moiré spin lattices within our linearly optical system, these lattices exhibit a property known as slow light. The group velocity of a SPP plane wave in air is given by $v_{gp} = kc/\beta < c$[38]. In contrast, for certain photonic Moiré spin lattices, the group velocity $v_g = |\mathbf{v}_g| = |\mathbf{P}/W|$ - defined as the ratio of the Poynting vector $\mathbf{P}$ (which is proportional to the kinetic Poynting momentum $\mathbf{p} = \mathbf{P}/c^2$)[38] to the energy density $W$ - can be significantly lower than the group velocity of the SPP plane wave, $v_{gp} = kc/\beta$.

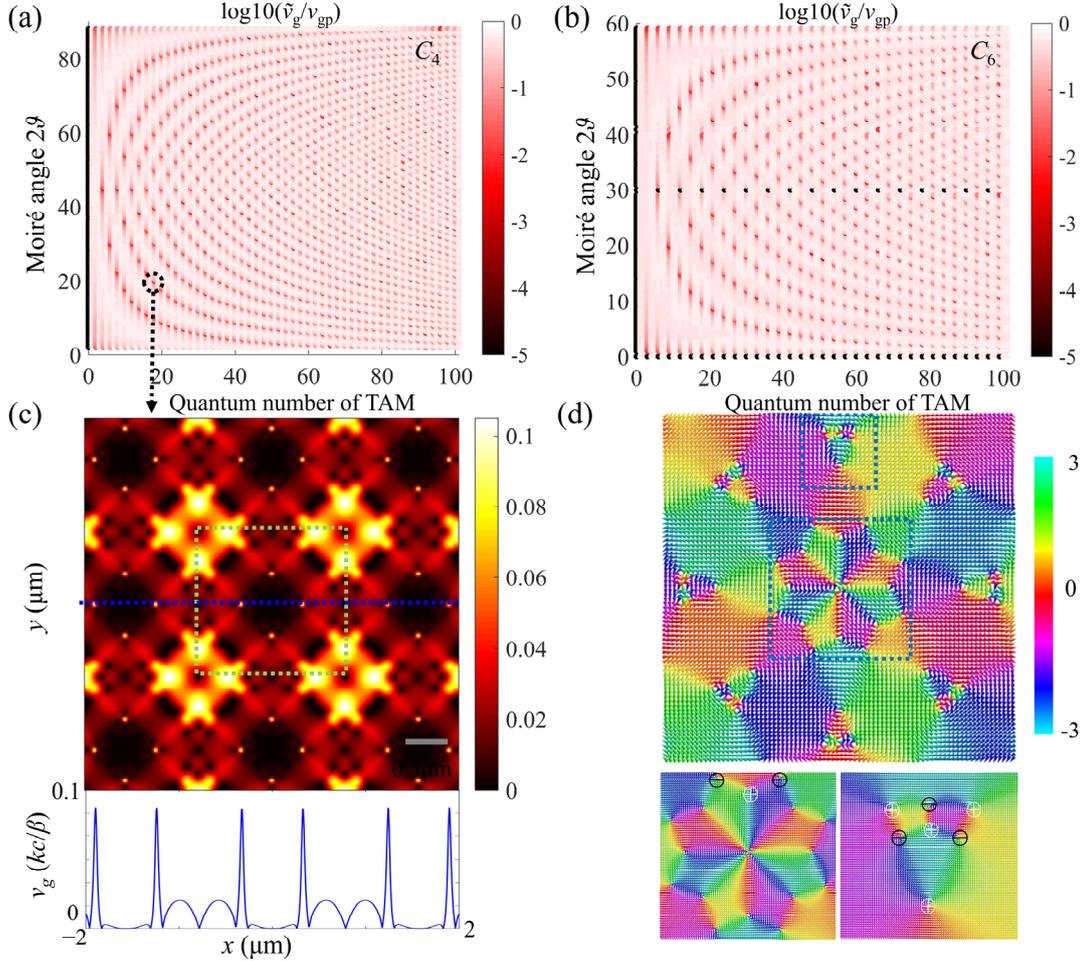

**FIG. 4. Slow light in photonic Moiré spin lattices.** The log10 of mean group velocity ($\log_{10}(\tilde{v}_g/v_{gp})$) is examined in relation to the Moiré angle $2\vartheta$ and the integer quantum number of TAM (from 0 to 101) for both (a) $C_4$ and (b) $C_6$ rotational symmetries. It is evident that in certain regions, the mean group velocity ($\tilde{v}_g$) is significantly lower (about several orders of magnitude smaller) than the group velocity of SPP in vacuum, $v_{gp} = kc/\beta$. For example, with a twisted angle of $2\vartheta_{C4} = \arctan(5/12)$ and the quantum number $l = 16$, the (c) 2D distribution of the local group velocity in the resulting photonic Moiré spin lattices is shown, along with its 1D contour along the $x$-axis (indicated by the blue dashed line in (c)). This analysis reveals that the maximum group velocity reaches about one-tenth of the $v_{gp}$. This phenomenon is closely related to the presence of off-axis vortex-antivortex flux and the optical super-oscillation effect, as illustrated in (d) and the enlarged figures in the bottom panel, the vector diagram of normalized group velocity within the square region surrounded by the green dashed line in (c) shows the spatial distribution of velocities. In this diagram, the red ⊕ symbol represents the vortex, while the blue ⊖ symbol denotes the antivortex. All velocities are normalized by $v_{gp}$ and the mean group velocity is calculated as $\tilde{v}_g = \Sigma(v_g \cdot W)/\Sigma W$ over a region of 10μm×10μm with a grid size of 0.005μm.

We present the calculated the $\log_{10}$ of mean group velocity ($\log_{10}(\tilde{v}_g/v_{gp})$) as a function of the Moiré angles described in Eq. 4 and Eq. 6, as well as the integer quantum number $l$. The results are shown in **FIG. 4**(a-b). For instance, in the Moiré lattice formed by the $C_4$ rotational symmetric sublattices with a twisted angle of $2\vartheta_{C4} = \arctan(5/12)$ and a TAM quantum number $l = 16$, the maximum local group velocity ($\tilde{v}_g$) of this Moiré lattice is found to be less than one-tenth of the $v_{gp}$, as depicted in the top panel of **FIG. 4**(c). The corresponding 1D contour can be seen in the bottom panel.

This slow light phenomenon arises due to the formation of local vortex-antivortex flux in the Moiré lattice (**FIG. 4**(d) and the enlarged figures in the bottom panel). This off-axis vortex-antivortex flux is intricately related to the optical super-oscillation effect[39,40] and can facilitate the creation of deep-subwavelength fine spin structures in confined optical fields[30]. Further details of the experimental results and their corresponding theoretical comparisons can be found in **FIG. S**21.

## Discussion

The integration of spintwistronics into photonics has significantly enhanced the flexibility in controlling topological spin superlattices. Our findings demonstrate that photonic bilayer spin superlattices can generate complex periodicities, novel tunable skyrmion topologies, fractal patterns, and facilitate slow light phenomena. Nonetheless, substantial opportunities remain for further exploration in designing spin superlattices with unique properties. For instance, studies of trilayer spin lattices[17], twisting layers with varied skyrmion topologies, and other innovative approaches inspired by advances in twistronics could further enrich this field. Moreover, introducing nonlinear effects into photonic spin lattices could emulate nonlinear interlayer coupling, potentially leading to the emergence of flat bands and the manifestation of van der Walls forces[11,12].

Beyond enabling advanced manipulations of electromagnetic fields with diverse topological configurations, the advent of optical spintwistronics is set to unlock a broad spectrum of applications. Photonic Moiré spin lattices, governed by spin-orbit couplings of light and intricately interacting with material chirality, offer promising opportunities in TAM-based optical data storage[41], chiral manipulations[42], atomic-scale chiral light localization[13], chiral sorting[43], and chiral laser emissions[44]. The precisely controllable photonic spin superlattices are poised to elevate optical tweezers[45] and quantum simulations using ultracold atoms via optical trapping technologies to higher-dimensional applications[47], and enhance super-resolution precision in optical sensing and metrology[40]. Additionally, the realization of diverse, particle-like topologies in optical spin superlattices holds great promise for high-density optical information storage and retrieval[46]. Beyond photonics, the methodologies developed in optical spintwistronics can be extended to encompass broader fields of spin physics, including acoustic spin[48], water-wave spin[45], and elastic wave spin[49], thereby pushing the boundaries of spintwistronics research even further.

**Acknowledgements:** This research was supported by the Guangdong Major Project of Basic and Applied Basic Research No. 2020B0301030009; National Natural Science Foundation of China (12174266, 92250304, 61935013, 12104318); Shenzhen Peacock Plan (KQTD20170330110444030); Scientific Instrument Developing Project of ShenZhen University (No.2023YQ001); Shenzhen University 2035 Initiative (No.2023B004); Nanyang Technological University Start Up Grant, Singapore Ministry of Education (MOE) AcRF Tier 1 grant (RG157/23), MoE AcRF Tier 1 Thematic grant (RT11/23), Imperial-Nanyang Technological University Collaboration Fund (INCF-2024-007).

**Author contributions:** P. S., Q. Z, and Y. S. conceived the basic idea of the work and supervised its development. P. S., X. G. and W. W. designed and performed the experiments. P. S., Q. Z, and Y. S. derived the theory, analyzed the experimental data and wrote the draft. All authors took part in discussions, interpretations of the results, and revisions of the manuscript. X. Y. supervised the project.

**Competing interests:** The authors declare no competing interests.

**Data and materials availability:** All data needed to evaluate the conclusions in the paper are present in the paper and the Supplementary Materials.

# Methods

In our work, the experimental setup used to demonstrate the photonic Moiré lattices is shown in **FIG. S**13(a), utilizing SPPs as the example. A linearly polarized beam with a wavelength of 632.8 nm, sourced from a He-Ne laser, served as the excitation source. The incident beam first passed through a linear polarizer (LP) and a half-wave plate (HWP) to achieve the desired linear polarization. Following this, the beam was modulated by a spatial light modulator (SLM), which controlled the phase of the beam according to the expression:

$$\psi = \arg\left[\sum_{n=1}^{N} e^{ik_t(x\cos(\theta_n+\vartheta)+y\sin(\theta_n+\vartheta))}e^{il(\theta_n+\vartheta)} + \sum_{n=1}^{N} e^{ik_t(x\cos(\theta_n-\vartheta)+y\sin(\theta_n-\vartheta))}e^{il(\theta_n-\vartheta)}\right] \quad (8)$$

to generate the desired pair of plane waves. The parameter $k_t$ was carefully chosen to match the beam size with the pupil of the objective lens. The integer $l$ was introduced to produce the desired vortex phase. The beams subsequently passed through a HWP and 1$^{st}$-order vortex-wave plate (VWP) to create radially polarized light. This light was then tightly focused using an oil-immersion objective lens (Olympus, NA=1.49, 100×), exciting the SPP plane waves on a gold film (50nm). The surface EM field was measured using an in-house near-field scanning optical microscopic system (NSOM). A polystyrene (PS) nanoparticle probe with a diameter of 300 nm was positioned on the gold film. The position of the PS particle was precisely controlled by a piezo-stage (Physik Instrumente, P-545). The near-field signal scattered by the PS particle was directed to an objective lens (Olympus, NA = 0.7, 60×), which acted as a low-pass filter. This signal was then split and analyzed using a combination of a QWP and a LP to extract the individual circular polarization component ($I_{LCP}$ and $I_{RCP}$) of the scattering signal. Using dipole theory, the horizontal components of the electric field could be reconstructed. These components were subsequently directed to two photomultiplier tubes (PMTs) for measuring the intensity information of the two signals. This setup enabled the characterization of the out-of-plane SAM component (i.e., along the optical axis) of the focused beams, as described by the relation[37]:

$$S_z = \frac{\varepsilon}{4\omega}\frac{\beta^2}{k_z^2}\left(I_{RCP} - I_{LCP}\right) \propto I_{RCP} - I_{LCP} \ . \quad (9)$$

The horizontal SAM component can be reconstructed by the Maxwell's EM theory[34,35], as introduced in Section V in Supplemental materials.